\documentclass[twoside]{article}
\usepackage{fleqn,espcrc2}

\usepackage{epsfig}
\usepackage[figuresright]{rotating}

\newcommand{\AmS}{{\protect\the\textfont2
  A\kern-.1667em\lower.5ex\hbox{M}\kern-.125emS}}

\title{The BAIKAL neutrino project: status report}

\author{
V.A.Balkanov\address{ Institute for Nuclear Research,
Moscow, Russia \\
$^b$ Irkutsk State University,
Irkutsk, Russia\\
$^c$ Institute of Nuclear Physics, MSU,
Moscow, Russia\\
$^d$ Nizhni  Novgorod  State  Technical University,
Nizhni  Novgorod, Russia\\
$^e$ St.Petersburg State  Marine Technical  University,
St.Petersburg, Russia\\
$^f$  Kurchatov Institute,
Moscow, Russia\\
$^g$ Joint Institute for Nuclear Research,
Dubna,Russia\\
$^h$ DESY-Zeuthen,
Zeuthen, Germany\\
$^i$ KFKI, Budapest, Hungary \\
\vspace {4mm}
presented by G.V.Domogatsky},
I.A.Belolaptikov$^g$, L.B.Bezrukov$^a$,
N.M.Budnev$^b$, A.G.Chensky$^b$, I.A.Danilchenko$^a$,
Zh.-A.M.Dzhilkibaev$^a$,
G.V.Domogatsky$^a$, A.A.Doroshenko$^a$, S.V.Fialkovsky$^d$,
O.N.Gaponenko$^a$, O.A.Gress$^b$, D.D.Kiss$^i$, 
A.M.Klabukov$^a$, A.I.Klimov$^f$, S.I.Klimushin$^a$,
A.P.Koshechkin$^a$, 
V.F.Kulepov$^d$, L.A.Kuzmichev$^c$, Vy.E.Kuznetzov$^a$,
J.Ljaudenskaite$^b$, B.K.Lubsandorzhiev$^a$,
M.B.Milenin$^d$, R.R.Mirgazov$^b$, N.I.Moseiko$^c$,
V.A.Netikov$^a$, E.A.Osipova$^c$, A.I.Panfilov$^a$, Yu.V.Parfenov$^b$,
L.V.Pankov$^b$,
A.A.Pavlov$^b$, E.N.Pliskovsky$^a$, P.G.Pokhil$^a$, V.A.Poleshuk$^a$,
E.G.Popova$^c$,
V.V.Prosin$^c$, 
M.I.Rozanov$^e$, V.Yu.Rubzov$^b$, Yu.A.Semenei$^b$,
I.A.Sokalski$^a$, CH.Spiering$^h$,
O.Streicher$^h$, B.A.Tarashansky$^b$, T.Thon$^h$, G.Toht$^i$,
R.V.Vasiljev$^a$, R.Wischnewski$^h$, I.V.Yashin$^c$, V.A.Zhukov$^a$.}

\vspace {-1cm}       
\begin{document}

\begin{abstract}
 We review the present status of the Baikal Neutrino Project
and present preliminary results of a search for upward 
going atmospheric neutrinos, WIMPs and magnetic monopoles 
obtained with the  detector {\it NT-200} during 1998. Also 
the results of a search for very high energy neutrinos
with partially completed detector in 1996  
are presented.

\vspace{1pc}
\end{abstract}

\maketitle


\vspace {-2cm}
\section{Detector and Site}

The Baikal Neutrino Telescope  is deployed in Lake 
Baikal, Siberia, 
\mbox{3.6 km} from shore at a depth of \mbox{1.1 km}. 
{\it NT-200}, the medium-term goal of the collaboration
\cite{APP}, was put into operation 
at April 6th, 1998 and consists of 192
optical modules (OMs). 
An umbrella-like frame carries  8 strings,
each with 24 pairwise arranged OMs.
Three underwater electrical cables and one optical cable connect the
detector with the shore station. 

The OMs are grouped in pairs along the strings. They contain 
37-cm diameter {\it QUASAR} - photo multipliers (PMs) 
which have been developed
specially for our project \cite{OM2,Project}. The two PMs of a
pair are switched in coincidence in order to suppress background
from bioluminescence and PM noise. A pair defines a {\it channel}. 

A {\it muon-trigger}
is formed by the requirement of \mbox{$\geq N$ {\it hits}}
(with {\it hit} referring to a channel) within \mbox{500 ns}.
$N$ is typically set to 
\mbox{3 or 4.} For  such  events, amplitude and time of all fired
channels are digitized and sent to shore. 
A separate {\em monopole trigger} system searches for clusters of
sequential hits in individual channels which are
characteristic for the passage of slowly moving, bright
objects like GUT monopoles.

Here we present preliminary results of analysis of data,
which were accumulated in the first 234 live days of {\it NT-200}
as well as results obtained from the analysis of data taken with
{\it NT-96}, the 1996 stage of the detector.

\section{Separation of fully reconstructed neutrino events}

The signature of neutrino induced events is a muon crossing the
detector from below. The reconstruction algorithm is based on the
assumption that the light radiated by the muons is emitted under
the Cherenkov angle with respect to the muon path. 
We don't take into account light scattering because the characteristic
distances for atmospheric neutrino induced muons detection do not 
exceed 1$\div$2 scattering lengths of light in Baikal water 
(mean scattering angle cosine $\approx$0.88)\cite{APP}.

The algorithm uses a single muon model to reconstruct
events. 
We apply procedure rejecting hits, which are very likely
due to dark current or water luminosity as well as hits
which are due to  
showers and have large time delays with respect to expected
hit times from the single muon Cherenkov light.

Determination of the muon trajectory is based on the minimization of a
$\chi^2$ function with respect to measured
and calculated times of hit channels.
As a result of the $\chi^2$ minimization we obtain the track parameters
($\theta$, $\phi$ and spatial coordinates). 

The reconstruction yields a fraction of about $4.6 \cdot 10^{-2}$ 
of events which are reconstructed as upward going with respect 
to whole event sample fulfilling the trigger condition
$\ge 6/3$ (at least 6 hits on at list 3 strings).
That is still far from a suppression factor 10$^{-6}$ necessary
for the depth of {\it NT-200}. To reject most of the 
wrongly reconstructed events we use the set of quality criteria.
If the event doesn't obey any of chosen criteria, it is rejected as 
wrongly reconstructed.  Different to {\it NT-96} \cite{APP2}, the 
neutrino selection algorithm for {\it NT-200} operates with 
trigger $\ge 7/3$.

For {\it NT-200} neutrino search, the following cuts are most 
effective:
(1) a traditional $\chi^2_t$ cut; (2) the minimum track length 
in the array; (3) the probability of non-fired channels not to be hit
and fired channels to be hit; (4) the correlation of measured
amplitudes to the amplitudes expected for reconstructed track;
(5) an amplitude $\chi^2_a$ defined similar to the time $\chi^2_t$;
(6) the correlation between measured hit times and vertical distances
of channels in array (see eq.1 below). 

The efficiency of the procedure and correctness of the MC background
estimation have been tested with a sample of $2.8 \cdot 10^6$
MC-generated atmospheric muons and with MC-generated upward going
muons due to atmospheric neutrinos.
\begin{table}[h]
\label{cuttable}
\begin{center}
\caption{
The fraction of events passing cuts for experimental and
MC background sample.}
\begin{tabular}{|c|c|c|} \hline
Applying &exper. & MC backgr.  \\
     cuts   & sample &  sample \\ \hline\hline
 $\theta > 90^o$ &$4.9 \cdot 10^{-2}$ & $4.6 \cdot 10^{-2}$\\  \hline
 $+$ ``soft'' cut(2) &$2.2 \cdot 10^{-2}$ & $2.0 \cdot 10^{-2}$\\  \hline
 $+$ ``soft'' cut(4) &$1.1 \cdot 10^{-2}$ & $1.1 \cdot 10^{-2}$\\  \hline
 $+$ ``soft'' cut(3) &$5.4 \cdot 10^{-3}$ & $6.3 \cdot 10^{-3}$ \\ \hline\hline
\end{tabular}
\end{center}
\vspace{-12mm}
\end{table}
None of MC background events has passed all cuts. 
Unfortunately, the restricted statistics of the MC background 
sample does not allow us to compare the behavior of MC 
background and experimental samples at all levels of tracks rejection.
To demonstrate the principal agreement between the action of
the cuts to experimental and and MC samples, we show in Table 1
the fraction of events passing cuts on the same variables
but with softer cut values.

Data taken with {\it NT-200} between 1998 April and 1999 February 
cover 234 days life time. For this period we got $5.3 \cdot 10^7$
events with trigger $\ge 6/3$. The set of above criteria was 
applied to this sample yielding 35 events which pass all of them. 
This number is in good agreement with 31 events expected 
from neutrino induced muons for this period. The reconstructed 
angular distribution for upward going muons from the experimental 
sample after all cuts is shown in Fig.1.
\begin{figure}
\vspace{-2mm}
\epsfig{file=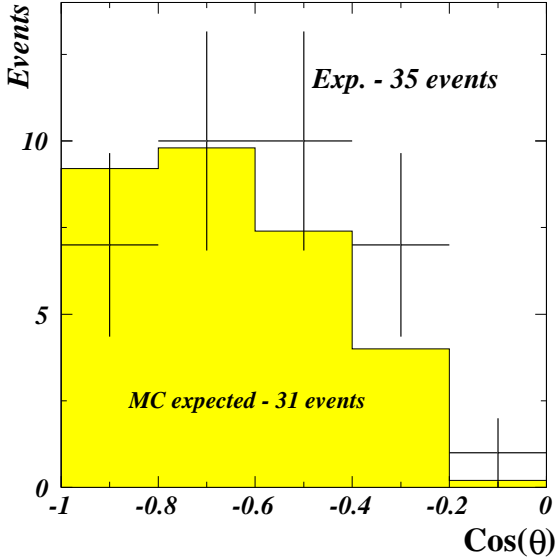,width=7.3cm}
\vspace{-0.8cm}
\caption{
Experimental angular distribution of reconstructed upward going 
muons in {\it NT-200}. Filled histogram - MC expected distribution.}
\vspace{-8mm}
\label{fig:angular3}
\end{figure}
 In the same figure the MC expected angular distribution
for muons from neutrinos is presented.

\section{Identification of nearly vertically upward moving muons}    

The search for weakly interacting massive particles (WIMPs) with the Baikal
neutrino telescope is based on the search for a statistically significant
excess of neutrino induced nearly vertically upward going muons
with respect to the expectation for atmospheric neutrinos.

Different to the standard analysis which has been described in
the previous section, the method of event selection relies on 
the application of a series of cuts which are tailored to the response
of the telescope to nearly vertically upward moving muons 
\cite{APP2,FRST_vert,JF}.
The cuts remove muon events far away from the opposite
zenith as well as  background events which are mostly due to
pair and bremsstrahlung showers below the array and to bare downward
moving atmospheric muons with zenith angles close to the horizon
($\theta>60^{\circ}$). The candidates identified by the cuts are 
afterwards fitted in order to determine their zenith angles.
%
\begin{figure}[htb]
\vspace{11pt}
\epsfig{file=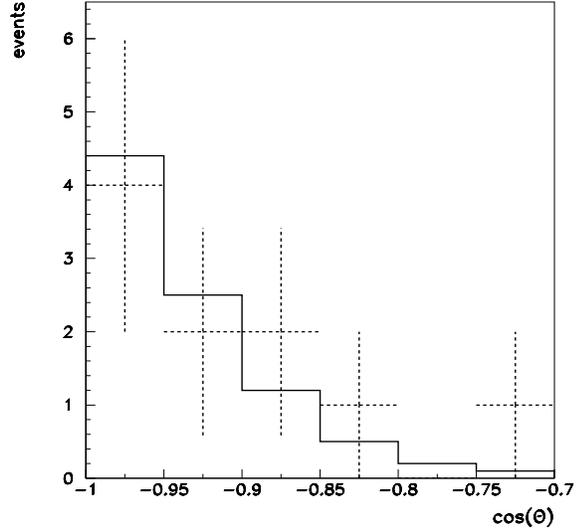,width=7.5cm,height=7.0cm}
\vspace{-1.2cm}
\caption { 
Zenith angular distribution of nearly vertically upward 
neutrino candidates as well as MC expectation for atmospheric
neutrino-induced muons (histogram).  
}
\label{fig2}
\vspace{-6mm}
\end{figure}

For the present analysis
we included all events with $\ge$6 hit channels, out of which 
$\ge$4 hits are along at least one of all hit strings.
To this sample, a series of 6 cuts is applied. Firstly,
the time differences of hit channels along each individual  string
have to be compatible with a particle close to the opposite zenith (1). 
The event length should be large enough (2), the maximum recorded
amplitude  should not exceed a certain value (3) and
amplitude of each upward looking hit channel has to be smaller then
a certain value (4).
The center of gravity of hit channels should not be close to
the detector bottom (5). The latter two cuts reject
efficiently brems showers from downward muons.
Finally, also time differences of hits along {\it different}
strings have to correspond to a nearly vertical muon (6).

The effective area of the full scale neutrino telescope {\it \mbox{NT-200}}
for muons with energy E$>$10 GeV, which move close to opposite zenith 
and fulfill all cuts, exceeds $2500$ m$^2$.

\begin{table}[h]
\vspace{-7mm}
\label{limit}
\footnotesize\rm
\caption{90\% c.l. upper limits on the muon flux from the center of the Earth
for six regions of zenith angles obtained in Baikal experiment}
\begin{center}
\begin{tabular}{||c|c|c|c||} \hline
Cone & Data & Back-  & Flux Limit \\
     &      & ground & (E$_{\mu}>$10GeV) \\
     &      & events & (cm$^{-2}$s$^{-1}$) \\ \hline
30$^{\circ}$  & 12 & 11.1 & 5.6$\times$10$^{-14}$   \\ \hline
25$^{\circ}$  & 9  & 9.1  & 4.0$\times$10$^{-14}$   \\ \hline
20$^{\circ}$  & 7  & 7.2  & 2.9$\times$10$^{-14}$   \\ \hline
15$^{\circ}$  & 4  & 4.4  & 2.0$\times$10$^{-14}$  \\ \hline 
10$^{\circ}$  & 2  & 1.5  & 2.4$\times$10$^{-14}$  \\ \hline 
5$^{\circ}$  & 1   & 0.5  & 1.7$\times$10$^{-14}$  \\ \hline \hline
\end{tabular} 
\end{center}
\vspace{-11mm}
\end{table}

From 234 days of effective data taking 
32957 events survive cut (1). 

After applying all cuts, ten events were selected as neutrino
candidates, compared to 8.9 expected from atmospheric neutrinos. 
The zenith angular distribution of these ten neutrino candidates 
is shown in fig.2.

Regarding the ten detected events as being due to
atmospheric neutrinos, one can derive an 
upper limit on the flux of muons from the center of the Earth   
due to annihilation of neutralinos - the favored candidate for
cold dark matter.

The combined numbers of observed and expected background events and 
the 90\% c.l.
muon flux limits for six cones around the nadir obtained with the Baikal
neutrino telescopes {\it NT-96} \cite{APP2} and {\it NT-200} (1998) are shown in Table 2. 

The comparison of Baikal flux limits with those obtained by 
Baksan \cite{Bak}, MACRO \cite{MACRO}, Kamiokande \cite{Kam} and
Super-Kamiokande \cite{SK} is shown in fig.3.

\begin{figure}[htb]
\vspace{-8mm}
\epsfig{file=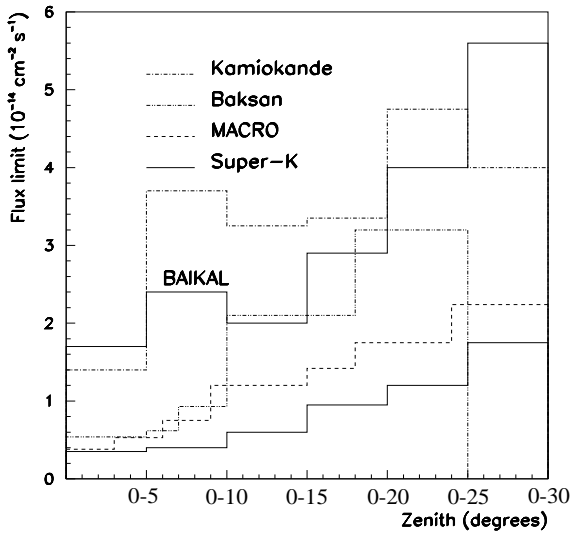,width=7.5cm,height=7.0cm}
\vspace{-12mm}
\caption { 
Comparison of Baikal nearly vertically upward muon flux limits
with those from other experiments.
}
\vspace{-8mm}
\label{fig3}
\end{figure}

\section{Search for fast monopoles ($\beta > 0.75$)}

Fast bare monopoles with unit magnetic Dirac charge and velocities greater
than the Cherenkov threshold in water ($\beta = v/c > 0.75$) are
promising survey objects for underwater neutrino telescopes. 
For a given velocity $\beta$ the monopole  Cherenkov  radiation exceeds that
of a relativistic muon by a factor $(gn/e)^2=8.3\cdot10^3$ ($n=1.33$ -
index of refraction for water) \cite{Fr,DA}.  
Therefore fast monopoles with $\beta \ge 0.8$ can be detected up to distances 
$55$ m $\div$ $85$ m corresponding to effective 
areas of (1$\div$3)$\cdot 10^4$ m$^2$.

The natural way to search for fast monopoles is based on the
selection of events with high multiplicity of hits
and high amplitudes. In order to reduce  the background from
downward atmospheric muons and especially atmospheric muon 
bundles we restrict ourself to monopoles coming from the lower hemisphere.

In the present analysis of the first 234 live days data of {\it NT-200},
the following cuts have been applied to the detected events.

\begin{itemize}

\item Number of hit channels $N_{hit} > 35$

\item The value of space-time correlation

\begin{equation}
cor_{zt}=\frac{1}{N_{hit}}\sum_{i=1}^{N_{hit}}
\frac{(z_i-\bar{z})(t_i-\bar{t})}{\sigma_{z} \sigma_{t}} > 0.6,
\end{equation}
where $z_i$ and $t_i$ are $z$-coordinate and time of hit channels,
$\bar{z}$, $\bar{t}$, $\sigma_z$ and $\sigma_t$ 
- their average values and standard deviations.

\item At least two of all hit channels have the amplitudes more than 400 ph.el.

\item The time differences of hit channels $\Delta t_{ij}$ fulfill the
following condition:

\begin{equation}
max(\Delta t_{ij} - \frac{R_{ij}}{v}) < 50 \mbox{ns},
\end{equation}
where $R_{ij}$ and $v$ - range between two hit channels and light
velocity in the water, respectively.

\end{itemize}
There are no events which survive all cuts. Using the MC calculated 
acceptance of {\it NT-200}, a 90\% c.l. upper limit on the monopole 
flux has been obtained.

The combined upper limit for an isotropic flux of bare fast magnetic monopoles
obtained with {\it NT-36}, {\it NT-96} \cite{BVEN} and
{\it NT-200} as well as limits from
underground experiments MACRO, Soudan2, KGF, Ohya and AMANDA 
\cite{MA,Sou,KGF,Oh,AVEN} are shown in Fig.4.

\begin{figure}[htb]
\epsfig{file=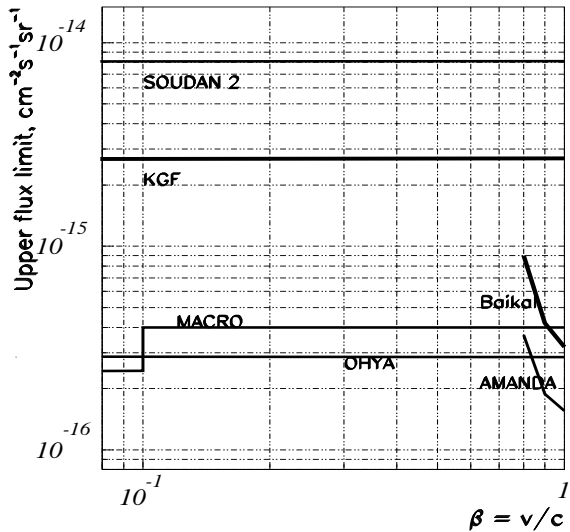,width=7.5cm,height=7.0cm}
\vspace{-15mm}
\caption { 
Upper limits on the flux of fast monopoles obtained in different
experiments.}
\vspace{-6mm}
\label{fig4}
\end{figure}

\section{The limit on the diffuse neutrino flux}

In this section we present results of a search for
neutrinos with $E_{\nu}>10 \,$TeV obtained with {\it NT-96} 
\cite{APP3}. 

The used search strategy for high energy neutrinos relies
on the detection of the Cherenkov 
light emitted by the electro-magnetic and (or) hadronic
particle cascades and high energy muons
produced at the neutrino interaction
vertex in a large volume around the neutrino telescope.

Within the 70 days of effective data taking of {\it NT-96}, 
$8.4 \cdot 10^7$ events with $N_{hit} \ge 4$ have been selected. 

For this analysis we used events with $\ge$4 hits along at least one
of all hit strings. The time difference between any two channels
on the same string was required to obey the condition:

\begin{equation}
\mid(t_i-t_j)-z_{ij}/c\mid<a\cdot z_{ij} + 2\delta, \,\,\, (i<j).
\end{equation}
The $t_i, \, t_j$ are the arrival times at channels $i,j$, and
$z_{ij}$ is their vertical distance. 
$\delta=5$ ns 
accounts for the timing error and $a=1 \,$ ns/m. 

8608 events survive the selection criterion (3).
The highest multiplicity of hit channels (one event) is $N_{hit}=24$.

Since no events with $N_{hit}>24$ are found in our data we can derive
an upper limit on the flux of high energy neutrinos which produce 
events with multiplicity N$_{hit}>$25.

The shape of the neutrino spectrum was assumed to behave like 
$E^{-2}$ as typically expected for Fermi acceleration.
In this case, 90\% of expected events would be produced by  neutrinos
from the energy range $10^4 \div 10^7$GeV.
Comparing the calculated rates with the upper limit to the 
number of zero events with $N_{hit}>24$,
we obtain the following  90\% c.l. upper
limit to the diffuse neutrino flux:

\begin{equation}
\frac{d\Phi_{\nu}}{dE}E^2<1.4\cdot10^{-5} 
\mbox{cm}^{-2}\mbox{s}^{-1}\mbox{sr}^{-1}\mbox{GeV}.
\end{equation}

\begin{figure}[htb]
\begin{center}
\vspace{-12mm}
\epsfig{file=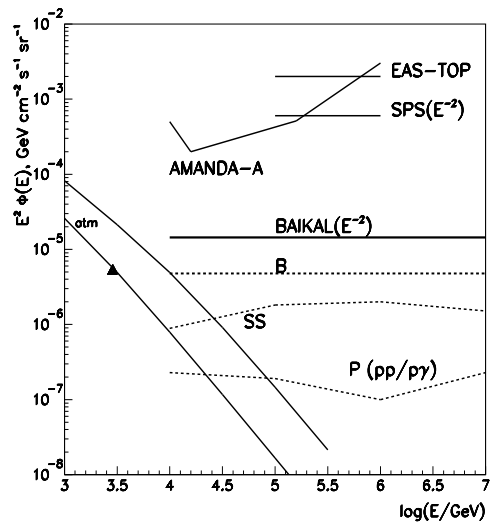,width=7.0cm,height=7.5cm}
\vspace{-11mm}
\caption { 
Upper limits on the differential flux of high 
energy neutrinos obtained by 
different experiments as well as upper bounds for  
neutrino fluxes from a number of different models. 
The triangle denotes the FREJUS limit.
\vspace{-6mm}
}
\end{center}
\label{fig5}
\end{figure}

Fig.5 shows the upper limits on the diffuse high energy neutrino
fluxes obtained by Baikal (this work), SPS-DUMAND \cite{DUMAND}, 
AMANDA-A \cite{AMANDA}, EAS-TOP \cite{EAS} and FREJUS \cite{FREJUS} 
(triangle) as well as a model independent upper limit 
obtained by V.Berezinsky \cite{Ber3} (curve labeled B)
(starting from the energy density of the diffuse X- and gamma-radiation 
$\omega_x \leq 2 \cdot 10^{-6}$ eV cm$^{-3}$
as follows from EGRET data \cite{EGRET}) and 
the atmospheric neutrino fluxes \cite{LIP} 
from horizontal  and vertical directions (upper and lower curves,
respectively). Also shown are  predictions from Stecker and Salamon 
model \cite{SS} (curve labeled SS) and Protheroe model \cite{P} 
(curve labeled P) for diffuse neutrino fluxes from quasar cores
and blazar jets.

We expect that the analysis of data taken with {\it NT-200} (1998) 
would allow us to lower this limit down to 
(2$\div$4)$\cdot$10$^{-6}$cm$^{-2}$s$^{-1}$sr$^{-1}$GeV. 

\section{EAS array and acoustic signal measurements}

Since March 1998 we have performed measurements 
of EAS with a Cherenkov array deployed on the ice cover 
just above {\it NT-200} \cite{BEAS}.

In March/April 2000 we continue the experiments with 
the EAS array. 
It consists of 4 upward facing {\it QUASAR} PMs  placed in 
special containers.
Three of them were located at the corners and one in the center
of an equilateral triangle. The distance between the central and each of 
the outer detectors was 100 m. The array was operating in 2 modes: 
a Cherenkov light detecting mode and a scintillator mode. For 
the Cherenkov mode 
conic reflectors were put on the containers to increase the effective
area of PMs. For the scintillator mode reflectors were replaced by 
 0.25 m$^2$ scintillator plates.

In the Cherenkov mode, the EAS array operated in coincidence with 
{\it NT-200} for studying the angular resolution of the latter. 
The energy threshold in this case was about 200 TeV. 
The preliminary analysis of data collected during 1999 shows that 
the angular resolution of {\it NT-200} \mbox{(without} applying any cuts, which
usually used to reject badly reconstructed tracks) is better than 5 degrees.
 
In the scintillator mode, the EAS array has been used as a trigger system 
in a search for acoustic signals from EAS. The core of EAS triggered
the scintillater array is expected to lead to an acoustic signal in
the ice and in the upper water layer. With 5 PeV energy threshold of 
the EAS-array, 2-3 events per hour have been observed. Acoustic 
hydrophone was placed 90 m apart from the center of the EAS array 
at a depth 5 m . 
Characteristic bipolar acoustic signals with about 150 $\mu$s 
duration and with a reasonable delay time compared to the EAS
trigger have been detected.
A preliminary analysis of the data shows that the amplitudes of the 
acoustic signals are somewhat larger than it would be expected from 
standard thermoacoustic theory \cite{LEAR}. The source of this
disagreement may be a rough calibration of hydrophon. 
We plan to continue the investigation of acoustic signals
from EAS in the next year.

\section{Conclusions and Outlook}

The deep underwater neutrino telescope {\it NT-200} in Lake Baikal is 
taking data since April 1998.
Using the first 234 live days, 35 neutrino induced upward muons
have been reconstructed. Although in a good agreement with MC
expectation this number is on factor 3 lower then predicted 
for the fully operational {\it NT-200}. The reason is that,  due to unstable
operation of electronics, in average only 50 - 70 channels have taken data
 during 1998. This is in contrast to 1999 and 2000 data taking where
stability had improved.
Ten events within a 30 degree half angle
cone around nadir have been selected and limits on the excess of muon flux
due to WIMP annihilation in the center of the Earth have been
derived. Also a new limit on the flux of fast monopoles has been obtained.

In the following years, {\it NT-200} will be operated as a
neutrino telescope with an effective area between 
1000 and 5000 m$^2$, depending on the energy.
It will investigate atmospheric neutrino spectra above 10 GeV
(about 1 atmospheric neutrino per two-three days).
Due to the high water transparency and low light scattering
with effective scattering length greater than 150m$\div$200m, 
the effective volume of {\it NT-200} for high energy electron and tau
neutrinos detection is more than two orders of magnitude
larger than its geometrical volume. This will permit
a search for diffuse neutrino fluxes from AGN 
and other extraterrestrial sources
on a level of theoretical predictions.

With an effective area two times larger than Super-Kamiokande,
for nearly vertically upward muons (E$_{\mu}>$10 GeV)  
{\it NT-200} will be one of the most powerful arrays for
 indirect search for WIMP annihilation in the center
of the Earth during the next few years.
It will also be a unique environmental laboratory to study 
water processes in Lake Baikal.

Apart from its own goals, {\it NT-200} is regarded to be a 
prototype for the development a telescope of next generation 
with an effective area of 50,000 to 100,000 m$^2$.
The basic design of such a detector is under 
discussion at present.

\bigskip

{\it This work was supported by the Russian Ministry of Research,the German 
Ministry of Education and Research and the Russian Fund of Fundamental 
Research ( grants } \mbox{\sf 99-02-18373a}, \mbox{\sf 00-02-31001} 
{\it and} \mbox{\sf 00-15-96794}) 
{\it and by the Russian Federal Program ``Integration'' (project no.} 346).

\end{document}